# Perpendicular Magnetization from Crystallite Boundaries in Polycrystalline Antiferromagnetic $\gamma$-FeMn

Rachel E. Maizel[1]*, Sachli Abdizadeh[1], Ali Almahmoud[1], Omolara A. Bakare[1], Galen T. Street[1], Hanu Arava[2], Sauviz P. Alaei[3,4], Sanyum Channa[3,4], Daisy A. O'Mahoney[4,5], Rishav Khatiwada[1], Jing Zhao[6], Charis Horn[7], Evan Stocker[1], Benjamin Kostroun[1], Emilio Corrales[1], Christoph Klewe[8], Alpha T. N'Diaye[8], Yuri Suzuki[4,9], Jean J. Heremans[1], Frederick M. Michel[6], Mitsuhiro Murayama[10,11], Vsevolod M. Ivanov[1,12,13], Satoru Emori[1]**

1. Department of Physics, Virginia Tech, Blacksburg, Virginia, USA
2. Materials Science Division, Argonne National Laboratory, Lemont, Illinois, USA
3. Department of Physics, Stanford University, Stanford, California, USA
4. Geballe Laboratory for Advanced Materials, Stanford University, Stanford, California, USA
5. Department of Materials Science and Engineering, Stanford University, Stanford, California, USA
6. Department of Geosciences, Virginia Tech, Blacksburg, Virginia, USA
7. National Center for Earth and Environmental Nanotechnology Infrastructure, Virginia Tech, Blacksburg, Virginia, USA
8. Advanced Light Source, Lawrence Berkeley National Laboratory, Berkeley, California, USA
9. Department of Applied Physics, Stanford University, Stanford, California, USA
10. Department of Materials Science and Engineering, Virginia Tech, Blacksburg, Virginia, USA
11. Institute for Materials Chemistry and Engineering, Kyushu University, Fukuoka, Japan
12. National Security Institute, Virginia Tech, Blacksburg, Virginia, USA
13. Virginia Tech Center for Quantum Information Science and Engineering, Blacksburg, Virginia, USA

*maizer@vt.edu; **semori@vt.edu

## Abstract

Single-crystalline noncollinear antiferromagnets, such as $Mn_3Sn$, are actively explored as memory media, but they generally require high-temperature growth that is incompatible with industrial-scale device fabrication. Here, we show that polycrystalline films of antiferromagnetic $\gamma$-FeMn, grown at room temperature, develop a perpendicular magnetization and an anomalous Hall effect, opening an industrially compatible route to antiferromagnetic memory. While conventional noncollinear antiferromagnets demand near-perfect crystal order, $\gamma$-FeMn films derive their memory-enabling properties from crystal defects, particularly boundaries separating nanoscale crystallites. When the crystallite size $D$ falls below ≈ 30 nm, a field-switchable magnetization emerges and increases with the boundary density $\propto 1/D$. This magnetization is accompanied by a room-temperature anomalous Hall conductivity exceeding 10 S/cm, comparable to that of single-crystalline $Mn_3Sn$. Our systematic experimental results, including scanning magnetometry, indicate uncompensated spin moments in the boundaries coupled to the compensated antiferromagnetism within the crystallites. We further find that the perpendicular easy axis of the magnetization is stabilized by residual strain within the $\gamma$-FeMn film. Our findings in manufacturing-friendly polycrystalline $\gamma$-FeMn redefine crystal defects as enablers, rather than impediments, for scalable antiferromagnetic memory.

## I. Introduction

The rapid advances of cloud computing and artificial intelligence present a critical need for fast, scalable magnetic memory devices[1]. They require thin-film materials with two essential properties: a switchable magnetic state along an easy axis, preferably perpendicular to the film plane, and an electrical readout of that magnetic state. Noncollinear antiferromagnets, such as $Mn_3X$ (X = Sn, Ge, Ga, Ir, Pt), combine both functionalities, while potentially offering orders-of-magnitude faster switching than conventional ferromagnetic media[2–5]. In noncollinear antiferromagnetic films, canted spin ordering can yield an uncompensated net magnetization with perpendicular anisotropy (an out-of-plane easy axis) that permits field- or current-driven switching[6,7]. At the same time, their distinctive crystal and spin structures can produce a large intrinsic anomalous Hall effect of topological origin[2,6,8,9], enabling electrical readout.

The scalability of these noncollinear antiferromagnets, however, poses a major problem. Their anomalous Hall effect relies on the long-range translational symmetry of the crystal, which chemical or structural disorder disrupts. Preserving the anomalous Hall effect has therefore typically required high-quality epitaxial growth on specialized substrates at temperatures exceeding 400°C[4–6,10–12], which are incompatible with complementary metal-oxide-semiconductor (CMOS) back-end-of-line integration[13,14]. Such stringent growth requirements remain a significant obstacle to scalable implementation of noncollinear antiferromagnetic memory.

Here, we show that microstructural defects, not long-range crystal order, give rise to both a perpendicular magnetization and an anomalous Hall effect in polycrystalline films of random-alloy, antiferromagnetic $\gamma$-FeMn. This alloy is thermally compatible with CMOS device fabrication and has long been used for exchange-biasing magnetic sensors[15–17]. Our work goes beyond the conventional use of $\gamma$-FeMn as a passive exchange-bias layer by demonstrating its potential as an active antiferromagnetic memory material. The $\gamma$-FeMn films here are sputter-deposited on Si/$SiO_2$ substrates with simple Ti/Cu seed layers, with no need for high-temperature processing. The resulting polycrystalline films exhibit a field-switchable nonzero perpendicular magnetization of < 0.1 $\mu_B$/atom and an anomalous Hall conductivity of > 10 S/cm at room temperature, comparable to those of highly crystalline noncollinear antiferromagnetic $Mn_3Sn$[6,9]. The magnetization and Hall effect emerge in the $\gamma$-FeMn films when the crystallite size $D$ is below ≈ 30 nm and increase with $1/D$. Our results indicate that crystallite boundaries, where translational crystal symmetry is broken, host uncompensated spin moments coupled to the antiferromagnetic crystallite interiors. These findings show that memory-enabling properties can emerge in polycrystalline antiferromagnetic media, whose behavior is governed by microstructural defects. Our work thus presents a scalable path to antiferromagnetic memory, where crystal imperfection is a resource, not a liability.

## II. Emergent Perpendicular Magnetization and Anomalous Hall Effect

FeMn alloy films were deposited by DC magnetron sputtering on Si substrates with a 50-nm-thick amorphous $SiO_2$ overlayer, unless noted otherwise. While $Fe_{50}Mn_{50}$ is the most commonly used composition[15–19], we opted for an $Fe_{60}Mn_{40}$ sputtering target. This composition lies near the eutectoid in the Fe-Mn phase diagram, where the face-centered cubic (FCC) $\gamma$ phase is thermodynamically stable closer to room temperature[20]. The FeMn films were seeded by either a Ti (3 nm) layer or a Ti (3 nm)/Cu (1-10 nm) bilayer. Unless noted otherwise, all film layers were grown and measured at room temperature.

Symmetric $2\theta$-ω X-ray diffraction (XRD) reveals that the seed layer strongly influences the crystal phase of FeMn films. For FeMn deposited directly on Ti [Fig. 1(a)], we observe a weak diffraction peak corresponding to the (110) plane of the body-centered cubic (BCC) phase of FeMn, also known as $\alpha$-FeMn[18,19]. In contrast, FeMn films grown on 1-nm Cu produce a pronounced diffraction peak from the (111) plane of FCC $\gamma$-FeMn [Fig. 1(b-d)], similar to textured growth of FCC metallic films in spin valves seeded by Cu[21,22]. We further corroborate the stabilization of FCC $\gamma$-FeMn on Cu with asymmetric grazing-incidence XRD and transmission electron microscopy (TEM) in the Supporting Information. Finally, exchange bias observed in Ni/$\gamma$-FeMn bilayers (Supporting Information) confirms that these $\gamma$-FeMn films are antiferromagnetic at room temperature as expected[15,17].

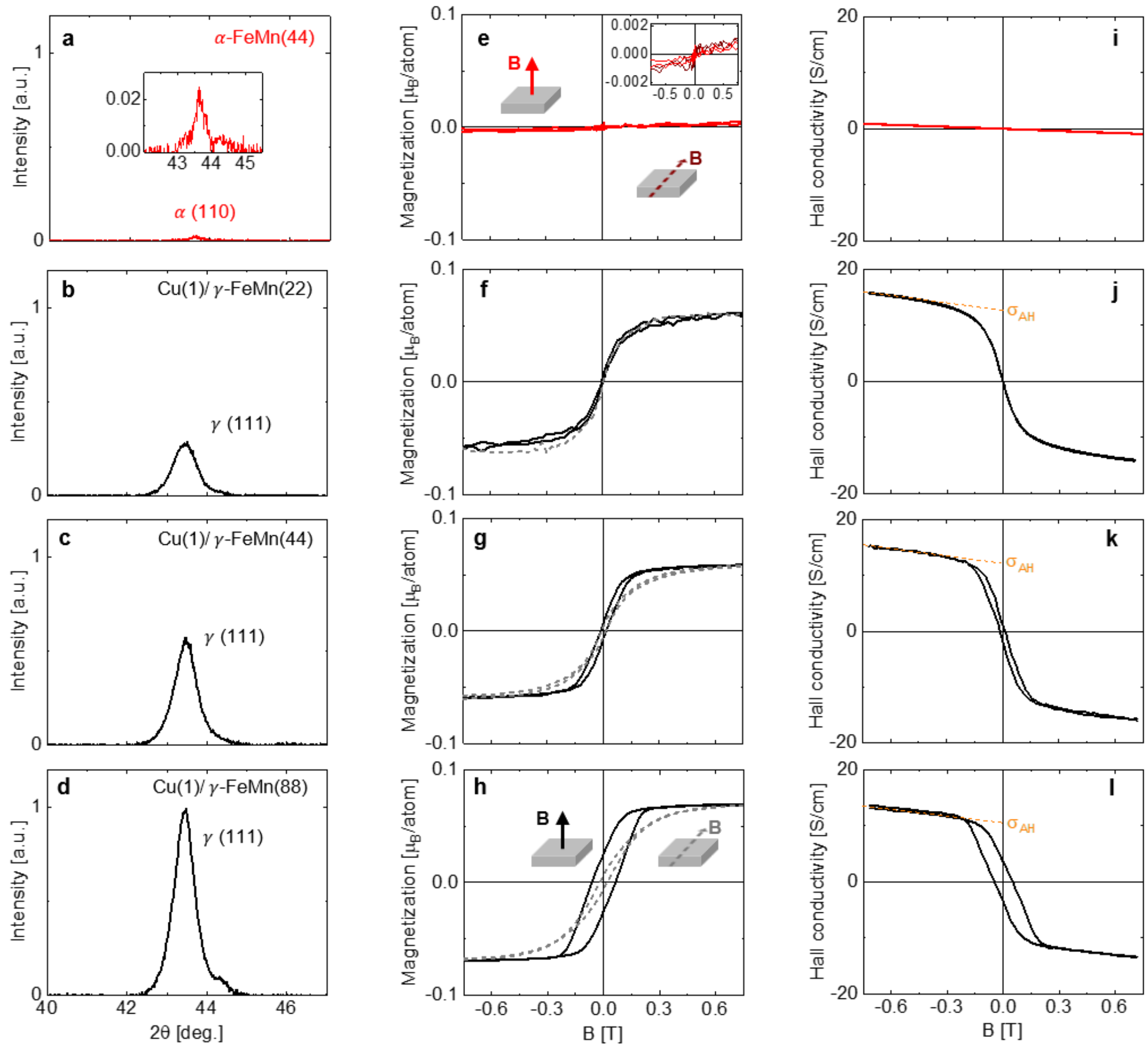

**FIGURE 1 |** (a) $2\theta$-ω XRD for a 44-nm-thick FeMn film with no Cu seed layer. (b-d) $2\theta$-ω XRD for 1-nm-Cu-seeded FeMn films with thicknesses (b) 22 nm, (c) 44 nm, and (d) 88 nm. The XRD intensity is normalized to the $\gamma$-FeMn (111) peak for Cu(1)/$\gamma$-FeMn(88). (e-h) In-plane (dashed curve) and out-of-plane (solid curve) magnetization curves for the same film stacks as (a-d). (i-l) Hall conductivity loops for the same film stacks as (a-d). In (j-l), the linear ordinary Hall background is retained; the dashed orange line marks the anomalous Hall conductivity at saturation $\sigma_{AH}$ extrapolated to $B$ = 0.

The $\alpha$ and $\gamma$ crystal phases of FeMn display qualitatively distinct magnetic behaviors. In $\alpha$-FeMn, no measurable magnetization is detected [Fig. 1(e)]. However, Cu-seeded $\gamma$-FeMn films exhibit a clear magnetic response with a volume-normalized saturation magnetization of $M_s \approx 0.06$-0.07 $\mu_B$/atom ($\approx$ 50 kA/m) [Fig. 1(f-h)]. This $M_s$ is more than an order of magnitude smaller than those of typical ferromagnetic metals and closer to the upper-bound values of antiferromagnetic $Mn_3X$ (~ 0.01 $\mu_B$/Mn atom)[6,10,23]. Remarkably, in $\gamma$-FeMn films thicker than 40 nm, the magnetization saturates at a lower field out-of-plane than in-plane. This out-of-plane easy axis implies uniaxial perpendicular magnetic anisotropy that overcomes the film-shape demagnetizing anisotropy of $K_d = \mu_0 M_s^2/2 \approx 2$ kJ/m$^3$. By integrating the area between the in-plane and out-of-plane magnetization curves for > 40-nm-thick $\gamma$-FeMn films, we find an effective perpendicular anisotropy energy density of $K_{u,eff} \approx 3$ kJ/m$^3$, corresponding to a uniaxial perpendicular anisotropy of $K_u = K_{u,eff} + K_d \approx 5$ kJ/m$^3$. The magnitude of $K_{u,eff}$ is comparable to that in noncollinear antiferromagnetic $Mn_3Sn$, which was recently incorporated into memory prototypes[4,24]. We also find that the magnetization in the $\gamma$-FeMn films persists up to at least 450 K (Supporting Information). Thus, Cu-seeded polycrystalline $\gamma$-FeMn films are antiferromagnetic media with a robust, switchable room-temperature perpendicular magnetization.

Hall measurements further reveal a significant difference between $\alpha$- and $\gamma$-FeMn films. In $\alpha$-FeMn, we observe only a small ordinary Hall effect that is linear with the field [Fig. 1(i)]. On the other hand, $\gamma$-FeMn films exhibit clear anomalous Hall hysteresis that mirrors the out-of-plane magnetization. For the samples in Fig. 1(j-l), we measure a room-temperature anomalous Hall conductivity of $\sigma_{AH} \approx 12$-14 S/cm, which is comparable to that in noncollinear antiferromagnetic $Mn_3Sn$[6,11,23]. We emphasize that the Cu-seeded polycrystalline $\gamma$-FeMn films attain the switchable perpendicular magnetization, along with the anomalous Hall response, on amorphous $SiO_2$ substrate surfaces without high-temperature processing. The emergence of the memory-enabling properties in $\gamma$-FeMn, grown at room temperature by industry-standard sputter deposition, makes it a highly attractive material for antiferromagnetic spintronics.

## III. Origin of the Perpendicular Magnetization

The observed magnetization is surprising, as strict spin compensation dictates zero magnetization in the symmetric antiferromagnetic $\gamma$-FeMn crystal. Broken symmetry is required to stabilize nonzero magnetization consisting of uncompensated spins, which must overcome the intrinsic antiferromagnetic exchange. Even more remarkable is the out-of-plane easy axis of the magnetization, which requires further symmetry breaking that overcomes the film-shape anisotropy.

One possibility is that uncompensated spins reside at the top and bottom surfaces of the $\gamma$-FeMn film. At the film surfaces, the lack of nearest neighbors disrupts the crystal translational symmetry and the antiferromagnetic exchange. In this case, the saturation magnetization $M_s$ and the effective perpendicular anisotropy $K_{u,eff}$, which are calculated as bulk quantities, should scale inversely with the film thickness $t$ [25]. In reality, we do not observe a systematic change in $M_s$ or $K_{u,eff}$ with $t$ [Fig. 2(a)]. The lack of inverse-$t$ scaling implies that the perpendicular magnetization originates predominantly from within the *volume* of the $\gamma$-FeMn film.

### A. Imperfect Antiferromagnetic Compensation at Crystallite Boundaries

The volume of a polycrystalline film consists of numerous nanoscale crystallites, separated by planar defects. Our cross-sectional TEM of a $\gamma$-FeMn film reveals columnar grains separated by grain boundaries [Fig. 2(f)]. Stacking faults can further subdivide a grain into crystallites stacked on top of each other as illustrated in Fig. 2(g). Thus, each crystallite is generally bounded by grain boundaries and stacking faults; henceforth, we refer to both as "boundaries."

While crystal translational symmetry is maintained within each crystallite, this symmetry is broken by the boundaries with reduced coordination. The broken symmetry disrupts the local antiferromagnetic exchange and precludes perfect spin compensation, hence giving rise to a net spin moment in the boundary. Compensated antiferromagnetism within ordered crystallites can coexist with uncompensated spins at defective boundaries, as is well established in studies of exchange bias[17,26–31]. The key picture is that uncompensated spins are largely confined to narrow, sub-nm-wide boundaries and, importantly, exchange-coupled to the crystallite's antiferromagnetic interior. $\gamma$-FeMn is qualitatively distinct from "weak ferromagnets," in which small net spin moments are distributed uniformly or surrounded by a nonmagnetic matrix. In this picture, the saturation magnetization $M_s$ in $\gamma$-FeMn is expected to rise with increasing density of boundaries or, equivalently, decreasing crystallite size.

We quantify the crystallite size $D$ (crystal coherence length perpendicular to the film plane) from the 2$\theta$-ω XRD peak widths of $\gamma$-FeMn, accounting for a microstrain (distribution of interplanar spacings) of 0.4% obtained from a Williamson-Hall analysis. Varying the $\gamma$-FeMn thickness $t$ > 40 nm and post-annealing at < 300°C do not appreciably affect $D$ [Fig. 2(c,d)]. In contrast, $D$ is significantly affected by post-annealing at > 300°C and varying the Cu seed thickness (1-10 nm). The latter tunes $D$ from ≈ 10 nm to ≈ 40 nm [Fig. 2(e)]; in thicker Cu seed layers, coalescence of Cu grains reduces the number of nucleation sites for $\gamma$-FeMn, resulting in larger crystallites.

Figure 2(e) plots $M_s$ against 1/$D$ (proportional to boundary density) and reveals crucial insights into the origin of the magnetization. The striking finding is that all samples with $K_{u,eff} > 0$ collapse onto one common trendline. (The one clear outlier in Fig. 2(e), "230°C anneal," has $K_{u,eff} < 0$ and is discussed further in the next subsection.) The trendline shows that nonzero $M_s$ emerges for crystallites smaller than the threshold of $D_0 \approx 30$ nm ($1/D_0 \approx 0.03$ nm$^{-1}$) and increases linearly with 1/$D$. To account for the 1/$D$ scaling[17,26–31], we model a crystallite as illustrated in Fig. 2(g): height $D$, width $D/\alpha$ ($\alpha \gtrsim 1$), boundary width $\delta \gtrsim 0.36$ nm (lattice parameter, $\delta \ll D$), and moment per boundary volume of $m_0$.  Following the derivation outlined in the Supporting Information, the trendline is expressed as

$$M_s = \delta(4\alpha+2)m_0(1/D - 1/D_0). \quad (1)$$

The slope, $\delta(4\alpha+2)m_0 \approx 1.5\ \mu_B$nm/atom in Fig. 2(e), corresponds to a boundary moment of $m_0 <$ 0.7 $\mu_B$/atom, below the atomic Fe or Mn moment of ≈ 1.5-2.0 $\mu_B$ in antiferromagnetic $\gamma$-FeMn[32–34]. This quantitative estimate implies that, for all $\gamma$-FeMn samples on the common trendline, the boundary spins are not *ferro*magnetically aligned but instead produce a net moment arising from *ferri*magnetic, canted, or frustrated spin order.

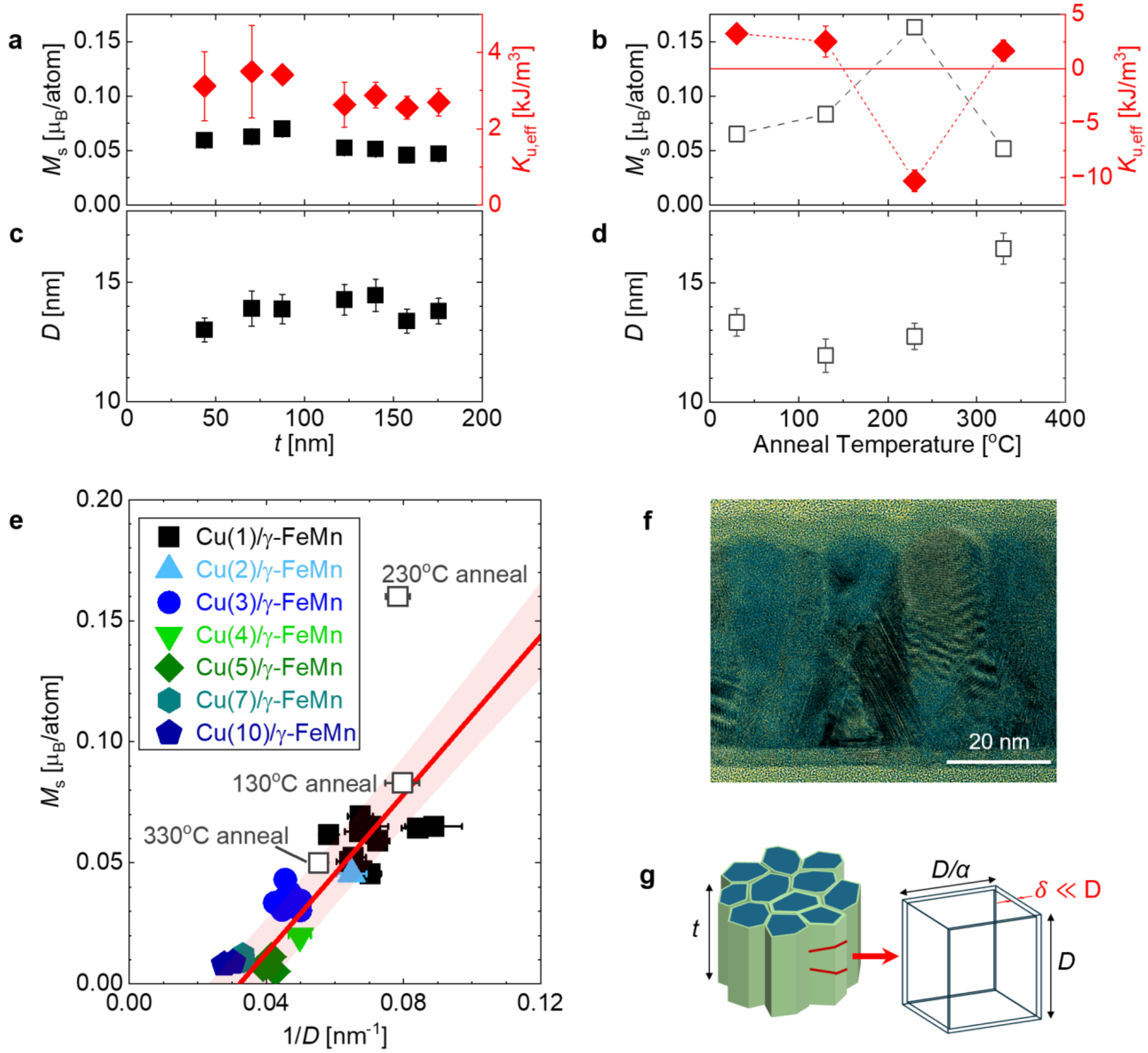

**FIGURE 2 |** (a) Saturation magnetization $M_s$ (black squares) and effective perpendicular magnetic anisotropy energy density $K_{u,eff}$ (red diamonds) of Cu(1)/$\gamma$-FeMn($t$) with thicknesses $t$ > 40 nm. (b) $M_s$ and $K_{u,eff}$ of Cu(1)/$\gamma$-FeMn(44) plotted against post-annealing temperature. (c,d) Crystallite size $D$ plotted against (c) $\gamma$-FeMn thickness $t$ and (d) post-annealing temperature. (e) $M_s$ vs 1/$D$ (~ crystallite boundary density) for Cu/$\gamma$-FeMn($t$ > 40 nm) with different Cu-seed layer thicknesses (1-10 nm) and for the same post-annealed samples as in (b,d). The linear fit (solid red line) is obtained from an ordinary least squares regression; the shaded region denotes the 68% (1σ) prediction band. (f) Cross-sectional transmission electron micrograph of Cu(1)/$\gamma$-FeMn(44). (g) Schematic of columnar grains, separated by grain boundaries (light green). Each columnar grain consists of stacked crystallites vertically divided by faults (red lines). A crystallite is approximated as a block of height $D$, width $D/\alpha$, and boundary width $\delta$.

The threshold $1/D_0$ in Fig. 2(e) is readily explained by a competition between two energy terms, analogous to the established energetics of exchange bias[29,35]: (1) the boundary surface energy favoring a common direction of the uncompensated ferrimagnetic-like moments and (2) the antiferromagnetic magnetocrystalline anisotropy energy barrier scaling with the crystallite volume. When the crystallites are too large and possess too high a magnetocrystalline energy barrier, the boundary moments (exchange-coupled to the antiferromagnetic crystallites) are frozen, unable to rotate to a common direction and cancelling out the globally averaged magnetization. When the boundary contribution surpasses the bulk crystallite contribution, the film develops a nonzero $M_s$ averaged over numerous boundaries. From this model, which we further explain in the Supporting Information, the threshold crystallite size $D_0$ is given by

$$D_0 = (4\alpha+2)\sigma_b/K_c \approx 30 \text{ nm}, \qquad (2)$$

where $\sigma_b$ is the boundary anisotropy energy density and $K_c$ is the bulk magnetocrystalline anisotropy energy density. Taking $K_c \approx 1.36\times10^5$ J/m$^3$ for $\gamma$-FeMn[30,36], we arrive at an upper bound of $\sigma_b < 0.7$ mJ/m$^2$, in line with exchange-biased $\gamma$-FeMn/ferromagnetic-metal bilayers[37]. Our model thus corroborates that the magnetization in polycrystalline $\gamma$-FeMn originates from crystallite boundaries exchange-coupled to the antiferromagnetic crystallite interior.

Furthermore, we conducted scanning nitrogen-vacancy (NV) magnetometry for real-space mapping of the magnetic stray field, with the spatial resolution set by the tip-to-sample standoff. Figure 3(a) shows the scanning NV results from an 80-nm-thick $\gamma$-FeMn film seeded with Cu, chosen for its large *average* crystallite size of $D \approx 40$ nm; the actual distribution of crystallite size is broad, allowing for even larger features that can be resolved by NV magnetometry. Although this $\gamma$-FeMn film has very little net saturation magnetization, we find localized stray-field features of order 1 mT throughout the scanned area. Both polarities occur in comparable numbers, as expected for local moments that are randomly oriented and cancelled on average. The observation confirms that even $\gamma$-FeMn with a globally averaged $M_s \approx 0$ still carries local moments on the scale of its crystallites.

More intriguingly, the largest features resolve into rings [Fig. 3(b)] in which an annulus of one polarity encloses the opposite polarity found within the ring. Our simulations [Fig. 3(c)] reveal that this ring-like structure is indeed consistent with an uncompensated moment surrounding an antiferromagnetic crystallite, whose diameter is ≈ 75 nm. Figure 3(d,e) further displays the vertical and horizontal line cuts of the target region in the experimental and simulated data, for which we observe a reasonable agreement. We note that the stray field lines close in on themselves, so a boundary magnetized in a single direction produces a field of one sign directly above it and of the opposite sign in the region it encloses. The reversed field inside the ring therefore does not require a second, oppositely magnetized domain. The asymmetric C shape of the ring is likewise a natural consequence of the NV-center geometry; since the NV axis is 54.78° from the film normal, the radially directed in-plane stray field adds to the measured projection on one side of the ring and subtracts from the other. Smaller ring-like structures are also likely present around smaller crystallites, but the spatial resolution of the current NV-center method precludes imaging of even finer structures. Nevertheless, our imaging results provide further support for the presence and critical role of uncompensated boundary spins in polycrystalline $\gamma$-FeMn films.

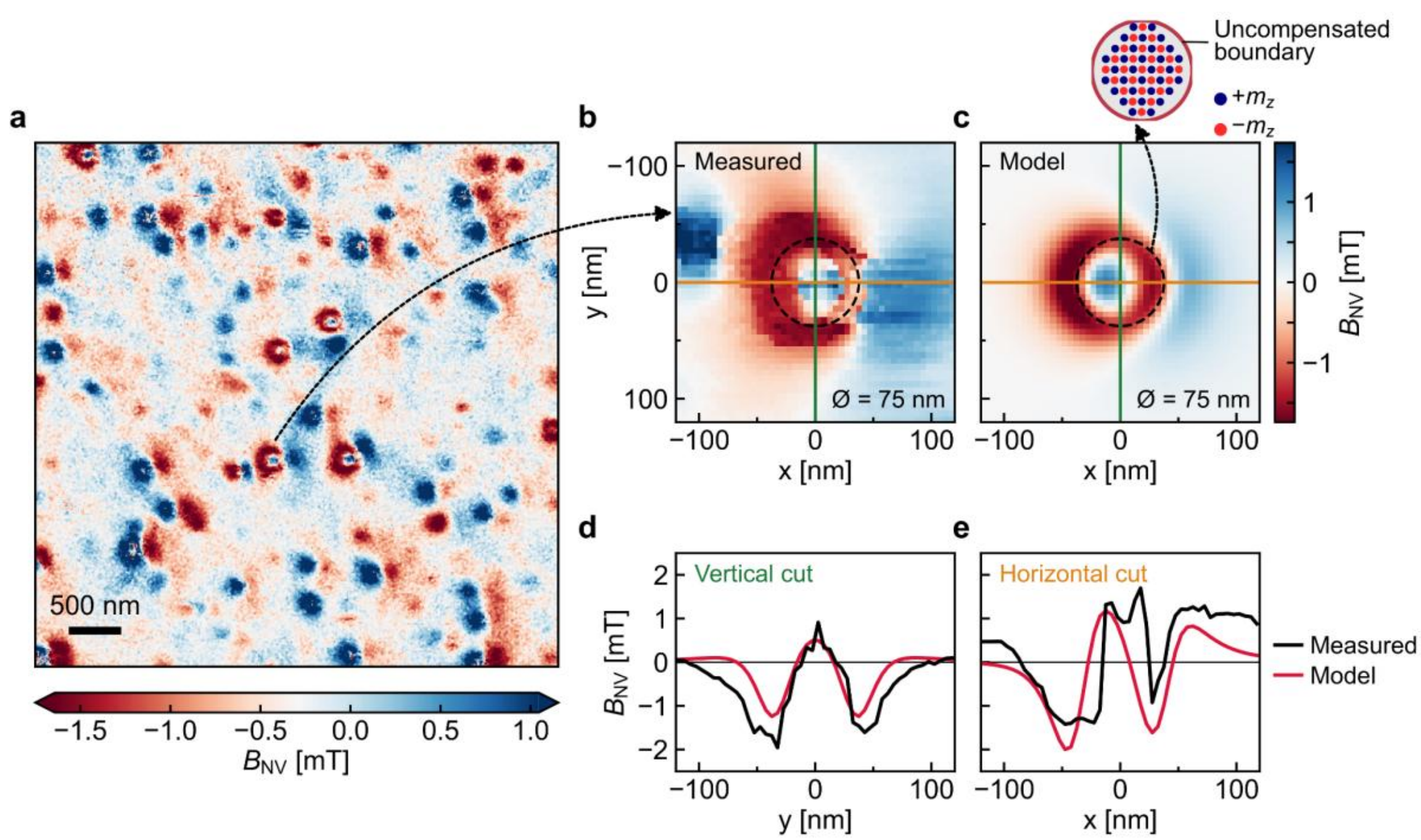

**FIGURE 3 |** Scanning NV-center magnetometry of polycrystalline $\gamma$-FeMn. (a) Stray field $B_{NV}$ above an 80-nm-thick $\gamma$-FeMn film seeded with 10 nm Cu. Despite $M_s \approx 0$ for the film as a whole, localized features of order 1 mT appear throughout in both polarities. (b) One of the largest features at higher resolution resolves into a ring. The dashed circle is 75 nm in diameter. (c) Stray field map simulated for an antiferromagnetic crystallite encircled by uncompensated boundary moments. (d) Vertical and (e) horizontal line cuts at the positions marked in (b) and (c) show a reasonable agreement between experimentally measured and simulated NV maps.

## B. Strain-Induced Perpendicular Easy Axis

While the crystallite-boundary-driven magnetism is of fundamental interest, the perpendicular anisotropy makes the antiferromagnetic $\gamma$-FeMn films particularly attractive for memory applications. The absence of the inverse-thickness scaling [Fig. 2(a)] already rules out a surface origin of perpendicular anisotropy. Instead, the most plausible origin is magnetoelastic anisotropy from global residual strain, which routinely arises *within* sputter-deposited films[38,39]. In as-grown $\gamma$-FeMn, the out-of-plane (111) interplanar spacing quantified from $2\theta$-ω XRD is 0.2083 ± 0.0006 nm, whereas the in-plane spacing from grazing-incidence XRD is 0.2059 ± 0.0018 nm. The lower in-plane spacing from grazing-incidence XRD is consistent with an in-plane compressive biaxial strain ($\varepsilon$ < 0). Given the reported large saturation magnetostriction of $\lambda_s$ ~ +100 ppm in antiferromagnetic FeMn[40,41], a small residual strain can produce magnetoelastic anisotropy $K_{me}$ that matches the observed uniaxial perpendicular anisotropy of $K_u$ ≈ +5 kJ/m³. Even with a conservative $\lambda_s$ of +20 ppm applied to $K_{me} = -3\lambda_s Y\varepsilon/2$ ($Y$ ≈ 200 GPa), the required in-plane biaxial strain is only $\varepsilon \approx -0.1\%$.

Thus, residual-strain-induced anisotropy can account entirely for the perpendicular anisotropy of the $\gamma$-FeMn films. We stress that the net magnetization is carried by uncompensated boundary spins, which are exchange-coupled to the antiferromagnetic crystallites [Sec. III A]. The global residual strain imprints a common magnetoelastic easy axis on the antiferromagnetic order of all crystallites; the boundary moments inherit this global easy axis through the exchange coupling. In contrast, magnetocrystalline anisotropy, governed by the crystallographic orientation in each crystallite, averages out to zero over numerous crystallites in the film.

Strain relaxation from mild post-annealing provides an additional test of the magnetoelastic interpretation. Upon post-annealing at 230°C [Fig. 2(b)], $K_{u,eff}$ experiences a sign reversal ($K_{u,eff}$ < 0) that corresponds to the easy axis changing from out-of-plane to in-plane, consistent with the removal of the strain-induced anisotropy. Even though $D$ remains essentially unchanged [Fig. 2(d)], $M_s$ more than doubles to 0.16 $\mu_B$/atom [Fig. 2(e)], equivalent to a boundary moment $m_0$ of up to ≈ 1.5 $\mu_B$/atom, suggesting a transformation of the magnetic order at the crystallite boundaries. Upon the loss of the strain-induced out-of-plane easy axis, the *ferri*magnetic, canted, or frustrated spin order in the boundaries appears to be transformed to *ferro*magnetic-like spin order.

Annealing at a higher temperature of 330°C coarsens the crystallites from $D \approx 13$ nm to 17 nm and restores $K_{u,eff} > 0$ [Fig. 2(b,d)]. This sample returns to the common $M_s$ vs $1/D$ trendline in Fig. 2(e). Within our magnetoelastic picture, the recovery of $K_{u,eff} > 0$ implies that in-plane compressive strain ($\varepsilon < 0$) re-emerges upon crystallite coarsening. Such compressive strain is not expected from grain growth alone; in substrate-clamped films, elimination of low-density boundary volume usually generates in-plane tensile stress[42,43]. Compressive strain can instead re-emerge through impurity segregation or the volume expansion associated with incipient BCC $\alpha$-phase formation[44]. Indeed, further increasing the annealing temperature to 430°C produces a predominantly BCC $\alpha$-Fe-rich phase. With this high-temperature annealing, $M_s$ rises sharply to 0.65 $\mu_B$/atom with easy-plane anisotropy, attributed to the dominance of $\alpha$-Fe ferromagnetism. A processing temperature below 400°C preserves the critical properties of polycrystalline antiferromagnetic $\gamma$-FeMn, along with CMOS-back-end-of-line compatibility. Unlike conventional noncollinear antiferromagnets that typically require crystallization above 400°C, polycrystalline $\gamma$-FeMn provides a low-thermal-budget medium for room-temperature antiferromagnetic memory.

## IV. Origin of the Anomalous Hall Effect

For the boundary magnetization to be useful in antiferromagnetic memory, it must be electrically readable. The anomalous Hall effect provides this readout pathway in $\gamma$-FeMn. To explore possible mechanisms, we examine the scaling of the anomalous Hall resistivity $\rho_{AH}$ with longitudinal resistivity $\rho_{xx}$. The scaling relation $\rho_{AH} \propto \rho_{xx}^n$ distinguishes skew scattering ($n \approx 1$) from the side-jump and intrinsic regimes ($n \approx 2$)[45,46]. As shown in Fig. 4(a), $\rho_{AH}$ scales approximately quadratically with $\rho_{xx}$, placing $\gamma$-FeMn in the dirty-metal regime and ruling out skew scattering.

This $\rho_{AH} \propto \rho_{xx}^2$ scaling can be consistent with a side-jump mechanism rooted in the boundary structure. The same crystallite boundaries that generate the uncompensated moments (time-reversal symmetry breaking) also introduce the spin-orbit-coupled disorder that mediates side-jump scattering[45]. Indeed, the anomalous Hall conductivity $\sigma_{AH}$ scales linearly with $M_s$ across all samples [Fig. 4(b)], linking the Hall response to the boundary magnetization. Nevertheless, precise disentanglement of side-jump from any residual intrinsic Berry-curvature contribution would require systematic transport measurements that vary both temperature and residual resistivity[47], which we identify as a target for future work.

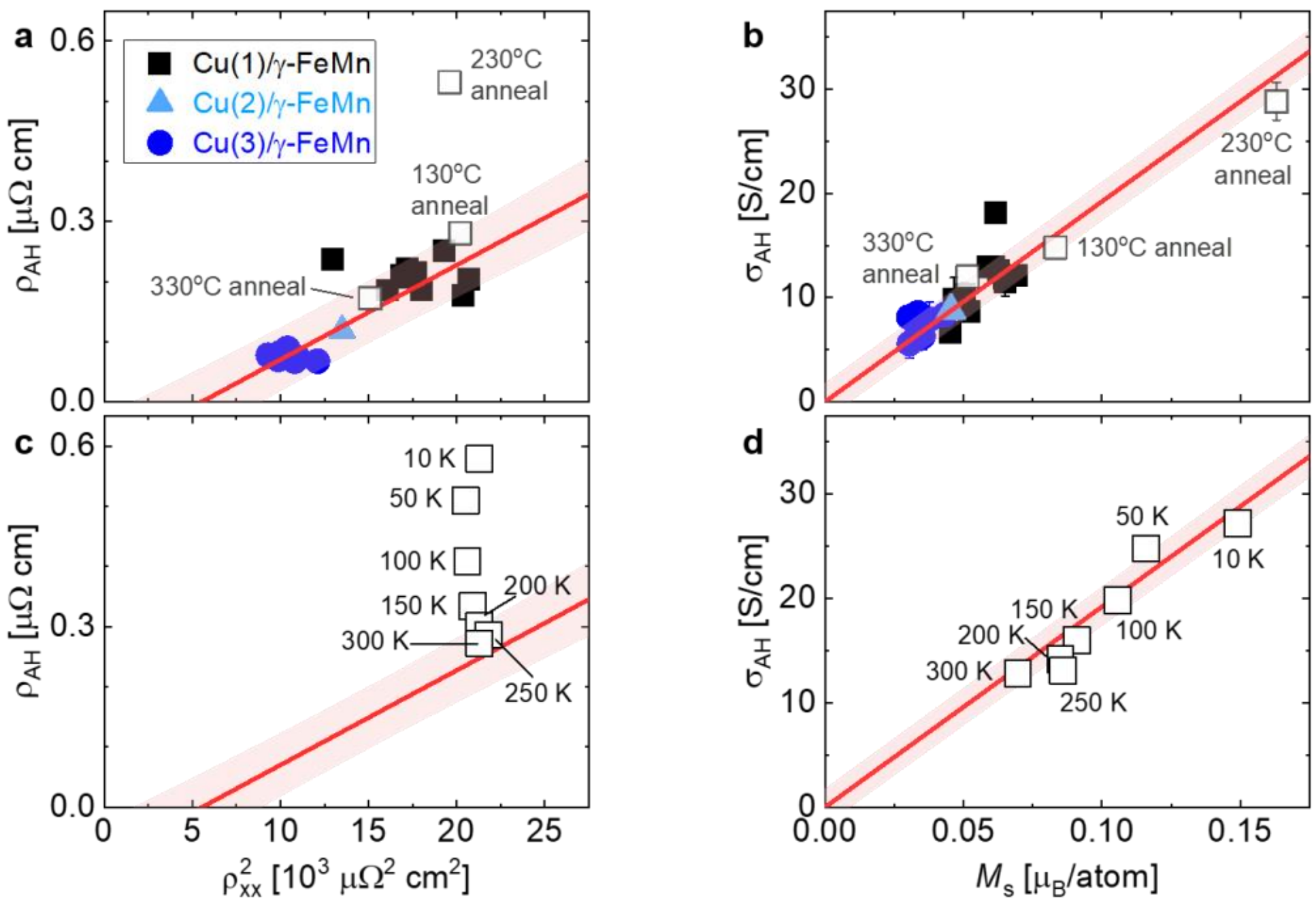

**FIGURE 4 |** (a) Anomalous Hall resistivity $\rho_{AH}$ as a function of longitudinal resistivity squared $\rho_{xx}^2$. (b) Anomalous Hall conductivity $\sigma_{AH}$ as a function of saturation magnetization $M_s$. The red lines are linear fits with shaded regions denoting the 68% (1σ) prediction bands. The maximum Cu-seed-layer thickness is limited to 3 nm; thicker Cu seed layers shunt an excessively large fraction of current away from $\gamma$-FeMn, making quantification of the transport parameters of $\gamma$-FeMn impractical. (c,d) $\rho_{AH}$ vs $\rho_{xx}^2$ and $\sigma_{AH}$ vs $M_s$ of Cu(1)/$\gamma$-FeMn(44) at different measurement temperatures. The red lines (bands) replicate the trendlines (prediction bands) from (a,b).

The 230°C-annealed sample clearly deviates from the common scaling of $\rho_{AH} \propto \rho_{xx}^2$ [Fig. 4(a)], even though this sample still falls on the same $\sigma_{AH}$ vs $M_s$ trendline as all other samples [Fig. 4(b)]. The deviation from $\rho_{AH} \propto \rho_{xx}^2$ suggests an additional mechanism governing the anomalous Hall effect, such as a change in magnetic order due to annealing-driven strain relaxation. Furthermore, the lower-temperature values of $\rho_{AH}$ of as-deposited $\gamma$-FeMn do not track the $\rho_{AH} \propto \rho_{xx}^2$ scaling at all [Fig. 4(c)], despite nearly following the linear $\sigma_{AH}$ vs $M_s$ scaling down to 10 K [Fig. 4(d)]. These results motivate further studies of the mechanisms underlying the anomalous Hall effect in $\gamma$-FeMn, which may help enhance electrical readout in antiferromagnetic memory.

## V. Ruling Out Competing Mechanisms

The findings above indicate that uncompensated crystallite-boundary spins give rise to the magnetization in the polycrystalline $\gamma$-FeMn films. Nevertheless, it is instructive to explore other possible mechanisms. For instance, given the high Fe content of the alloy, one alternative is that the magnetization originates from segregation of ferromagnetic BCC $\alpha$-Fe within the antiferromagnetic $\gamma$-FeMn matrix[18]. However, we do not observe a clear signature of such Fe segregation in energy-dispersive X-ray spectroscopy elemental maps [Fig. 5(a)]. We further compare the spectroscopic fingerprints of $\gamma$-FeMn and $\alpha$-Fe through *L*-edge X-ray magnetic circular dichroism (XMCD)[48], whose spectral lineshape is sensitively affected by the local bonding environment of the magnetic Fe atoms[49]. As displayed in Fig. 5(b), the XMCD spectrum of $\gamma$-FeMn exhibits a wider $L_3$ peak around 708 eV and an $L_2$ peak around 720 eV that is shifted and has a qualitatively distinct shape compared to the $\alpha$-Fe spectrum. These qualitative differences indicate that the Fe atoms carrying the magnetization of $\gamma$-FeMn do not reside in an $\alpha$-Fe-like environment. Thus, segregated $\alpha$-Fe cannot account for the observed magnetization.

Another possibility is the "topological orbital ferromagnetism" theoretically predicted in $\gamma$-FeMn with noncoplanar (3Q) spin order[50,51]. In essence, a sizable *orbital* magnetization, rooted in the real-space topology (scalar chirality) of the 3Q noncoplanar spins, could arise with uniaxial strain normal to the (111) crystal plane[50,51]. However, our calculations establish that the noncoplanar 3Q antiferromagnetic spin order is nearly degenerate with noncollinear 2Q and collinear 1Q antiferromagnetism (Supporting Information). The close degeneracy makes it difficult to uniformly stabilize the requisite 3Q spin order for topological orbital magnetism in the structurally disordered $\gamma$-FeMn films. Additionally, applying the sum-rule analysis[48] to $\gamma$-FeMn XMCD spectra, we find a small orbital moment of only ≈ 0.003 $\mu_B$/Fe atom, compared to the spin moment of ≈ 0.107 ± 0.003 $\mu_B$/Fe atom. The observed magnetization is most readily explained by uncompensated spin moments, not topological orbital moments, highlighting the critical role played by imperfections (i.e., crystallite boundaries) in the polycrystalline $\gamma$-FeMn films.

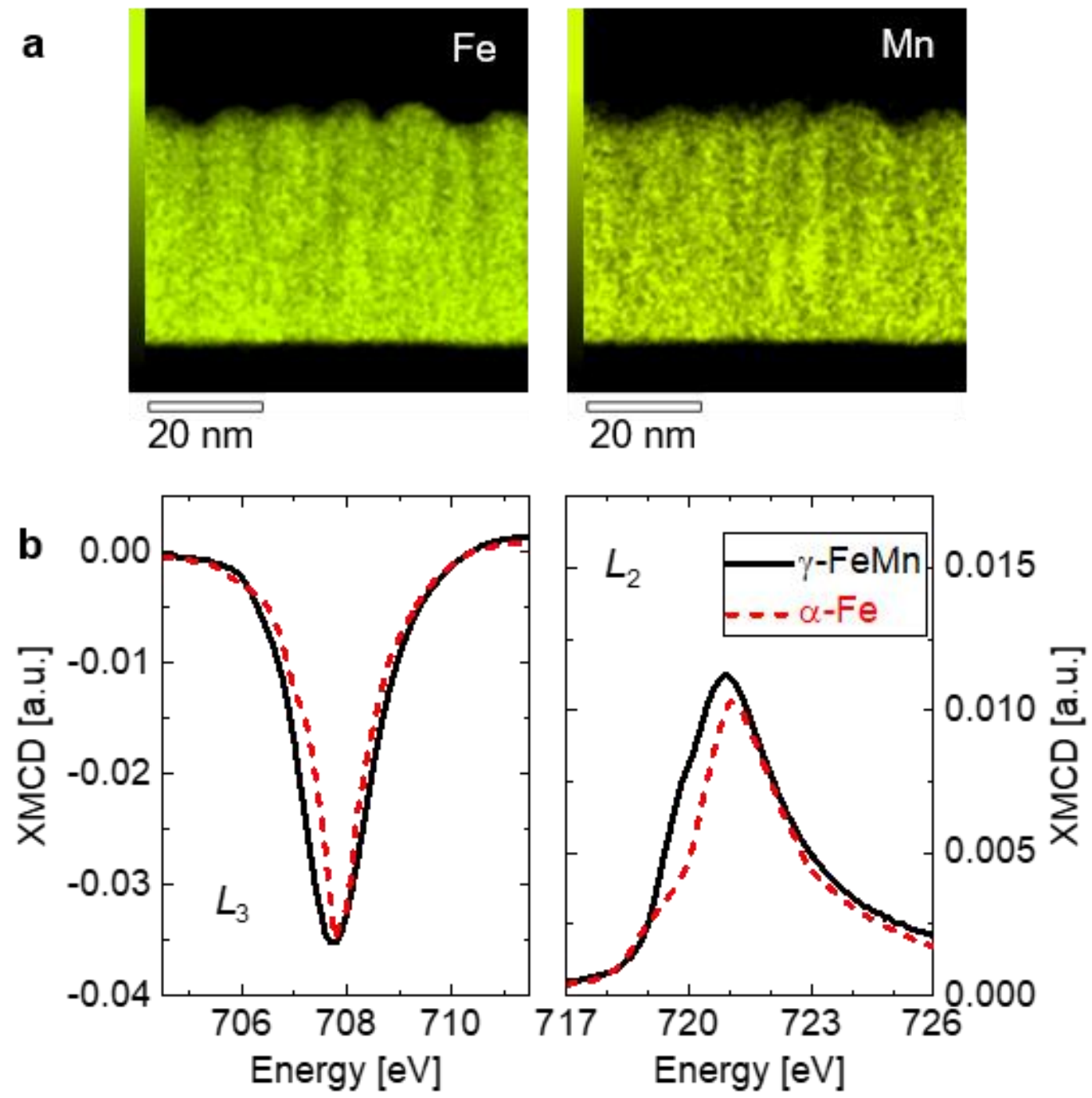

**FIGURE 5 |** (a) Cross-sectional energy dispersive X-ray spectroscopy elemental maps of Cu(1)/$\gamma$-FeMn(44), showing spatial distributions of Fe and Mn. (b) Fe *L*-edge XMCD spectra for 50-nm-thick $\alpha$-Fe and 44-nm-thick $\gamma$-FeMn, scaled such that their $L_3$-edge peaks are equal in amplitude.

## VI. Summary

We have demonstrated that structurally and chemically disordered polycrystalline films of antiferromagnetic $\gamma$-FeMn offer promising attributes for scalable memory. These $\gamma$-FeMn alloy films are easily grown by sputter deposition at room temperature on standard Si/$SiO_2$ wafers, in contrast to noncollinear antiferromagnetic compounds that generally require high-temperature growth to establish chemical and structural order. The $\gamma$-FeMn films exhibit a robust perpendicular net magnetization and an anomalous Hall conductivity exceeding 10 S/cm at room temperature, properties previously associated with highly crystalline noncollinear antiferromagnets. Rather than requiring a clean lattice, these properties in $\gamma$-FeMn are enabled by microstructural imperfections in the form of crystallite boundaries. This defect-driven paradigm provides an intriguing alternative to the conventional scheme in antiferromagnetic spintronics, in which near-ideal crystallinity has been a prerequisite. Our findings suggest that defect or grain-boundary engineering, a well-established approach in various materials technologies[52,53], offers a manufacturing-friendly route to antiferromagnetic memory media.

## Acknowledgements

R.E.M., A.A., O.A.B., and S.E. were supported by the National Science Foundation (NSF), Grant ECCS 2144333. S.A. and G.T.S. were supported by the National Science Foundation (NSF), Grant ECCS 2236160. S.P.A. and S.C. were supported by the Air Force Office of Scientific Research, Grant No. FA9550-23-1-0344. D.A.O. and Y.S. were supported by the U.S. Department of Energy, Director, Office of Science, Office of Basic Energy Sciences, Division of Materials Sciences and Engineering under Contract No. DE-SC0008505. H.A. was funded by the US Department of Energy, Office of Science, Office of Basic Energy Sciences, Materials Science and Engineering Division. This study used shared facilities at the Virginia Tech National Center for Earth and Environmental Nanotechnology Infrastructure (NanoEarth), a member of the National Nanotechnology Coordinated Infrastructure (NNCI), supported by the National Science Foundation (NSF), Grants ECCS 1542100 and ECCS 2025151. This research used resources of the Advanced Light Source, a U.S. DOE Office of Science User Facility under contract No. DE-AC02-05CH11231. E.S., B.K., E.C., and V.M.I acknowledge Advanced Research Computing (ARC) at Virginia Tech for providing computational resources and technical support. M.M. acknowledges financial support from the JSPS KAKENHI Grant No. 23H00238. During the preparation of this manuscript, the authors used Claude Opus 5, Kimi K3 (hosted by ARC), and GLM 5.3 (hosted by ARC) to assist with literature surveying and refining drafts; all literature citations and statements were reviewed and verified by the authors, who take full responsibility for the final contents of the publication.